\documentclass{article}
\usepackage{graphicx} 
\usepackage{float}
\usepackage{hyperref}
\usepackage{xcolor}
\usepackage{natbib}
\usepackage{color}
\usepackage{soul}
\usepackage[margin=3.5cm, footskip = 0.5in]{geometry}
\hypersetup{
    colorlinks=true,
    linkcolor=blue,
    filecolor=magenta,      
    urlcolor=cyan,
    citecolor=red
}

\title{Another look at predicting molecular breast cancer subtypes from the METABRIC data}
\author{Isabella Lazarov\thanks{ Email: il2482@columbia.edu}\\
    Columbia University \\
    and 
    \\
    Rob Tibshirani \\
    Departments of Biomedical Data Science and Statistics,\\Stanford University}

\date{June 2026}

\begin{document}

\maketitle

\begin{abstract}  
Classifying patients into different clusters based on genomic data can offer valuable insights into their projected disease-specific survival trajectories over time. Here we apply two supervised learning methods---Nearest Shrunken Centroids and LASSO--- to the METABRIC Breast Cancer data \citep{metabric} of 1980 patients and 754 genes to perform this task.  The {\tt pamr} R package implements the Nearest Shrunken Centroids classifier and the {\tt glmnet} R package is used to fit an ungrouped multinomial model, a grouped multinomial model, and a One-Versus-Rest model. Splitting our data into discovery and validation sets, we evaluate all four models' classification performance and the survival implications of their class predictions using cross validation and Kaplan-Meier curves. We find that the One-Versus-Rest model produces the lowest misclassification error of 0.0572 and the lowest median log-rank test of 0.380 statistic measuring the similarity between its Kaplan-Meier curves and the discovery set's true Kaplan-Meier curves. We show that a multinomial classification task split into several LASSO binomial classifiers offers promising results for patient clustering.
\end{abstract}

\section{Introduction}
Consider a high-dimensional dataset, each row representing a patient and each column measuring either RNA gene expression (GEX) or copy number variation (CNV). Several supervised models can be trained to classify the patients in this dataset into categories. By categorizing patients into groups representing different survival distributions, clinical researchers may use such models to predict a patient's potential survival trajectory and thus judge the necessity and appropriate strength of potential treatments.

\begin{figure}
    \centering
        \includegraphics[width=\textwidth]{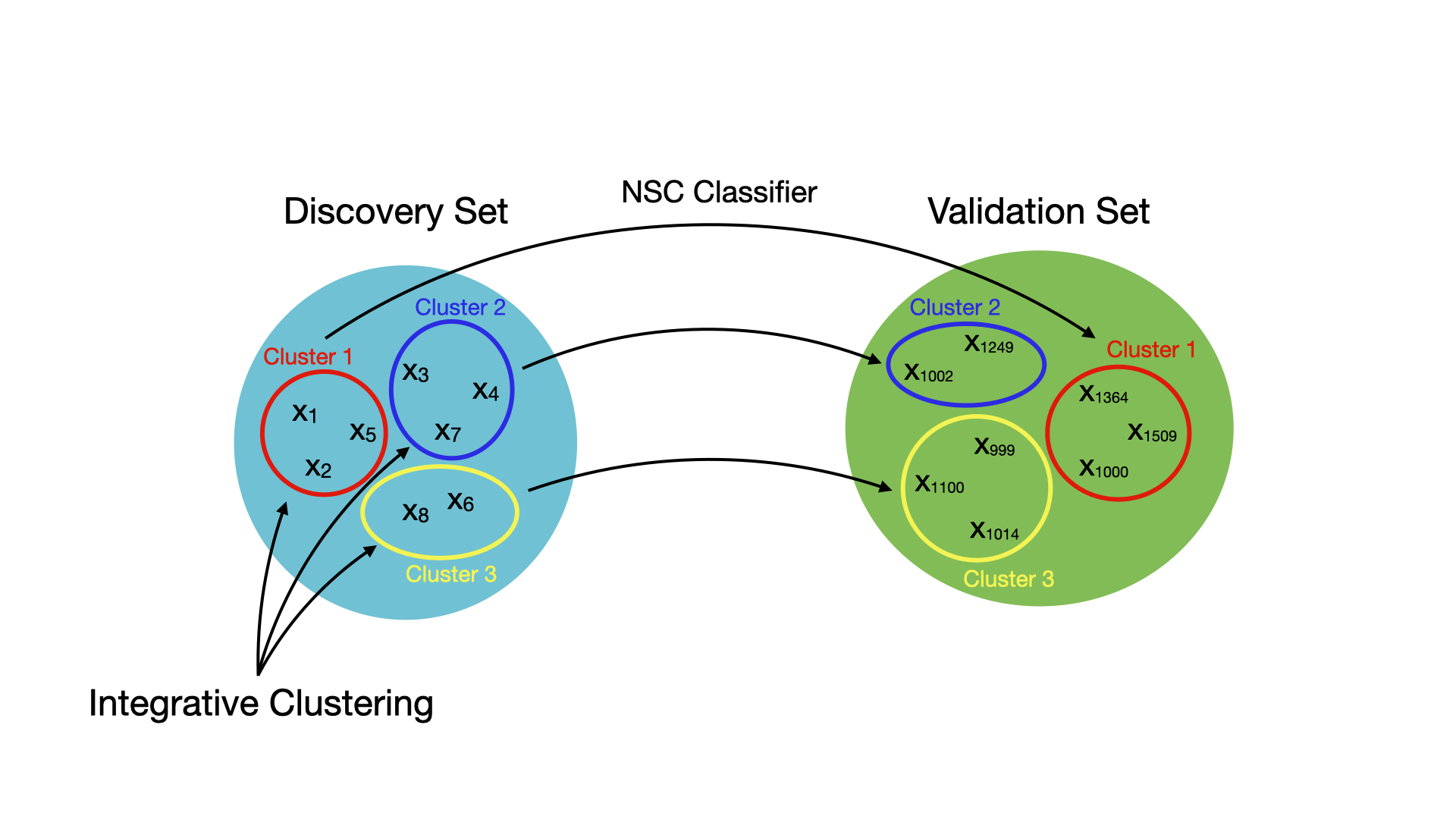}
        \caption{\textit{Graphic visualizing integrative clustering assignments for discovery set and NSC cluster classification of validation set. Integrative clustering analyzes the gene expression and copy number variation data of the discovery set and produces subgroups into which to divide its samples. The NSC classifier is trained on these ``true'' labels to predict the cluster subgroups of the validation set samples.}}
        \label{fig1}
\end{figure}

\citet{curtis2012} performed this categorization on the METABRIC breast cancer dataset which can be downloaded from the METABRIC cBioPortal page\footnote{\url{https://www.cbioportal.org/study/summary?id=brca_metabric}}. Splitting a dataset of 1980 rows into discovery and validation sets of 997 and 983 samples, respectively, they used an integrative clustering method to analyze the GEX and CNV data of the discovery set's breast cancers and divide them into different clusters. They then attempted to reproduce the cluster classification on the validation set using a Nearest Shrunken Centroids (NSC) classifier, whose reproducibility was affirmed in three ways: similar proportions of cases to the 10 clusters, similar hazard ratios between the discovery and validation sets, and the in-group proportions measure \citep{curtis2012}. A graphic of the classification method is shown in Figure \ref{fig1}.

As an alternative, we apply multiple LASSO classifiers to predict the cluster labels of the validation set. We evaluate their performance using cross validation (CV) error and we evaluate the similarity between their predicted clusters' final survival distributions and those of the discovery set using Kaplan-Meier (KM) curves and log-rank test statistics. For the classification task, we use the {\tt glmnet} R package to fit three different LASSO models: ungrouped multinomial, grouped multinomial, and One-Versus-Rest (OVR) binomial. We also reproduce the NSC classifier using the {\tt pamr} R package. From now on, we use ``{\tt pamr}'' and NSC interchangeably.

Note that this is an unusual classification task, in that we don't  have any ``ground truth'' labels in the validation set. We instead use the validation set survival data to assess prediction accuracy.

\section{Pipeline}
\subsection{Data Acquisition and Preprocessing}
As explained previously, we extracted the survival and cluster data from the \path{data_clinical_patient.txt} file from the cBioPortal page, removed any incomplete rows from both, and documented the ID numbers for the complete rows. The discovery set of 997 samples and 754 features and the validation set of 995 samples and 754 features, and their corresponding ID numbers, were downloaded from \citet[Supplementary Table S43]{curtis2012}. We did not remove any NA values in our gene data during data preprocessing as there is an imputation step right before cross validation later in the pipeline. We matched the sample IDs of the survival, cluster, and discovery gene data and then the IDs of the survival, cluster, and validation gene data. Twelve samples were lost in this ID matching step as seven patients in the validation gene data were absent from the survival data and the same seven patients along with another five were absent from the cluster data. Our preprocessing results in a final discovery set of 997 patients and a final validation set of 983 patients, both sets including their corresponding survival and cluster data. 

Since \citet{curtis2012} classified its validation set using NSC, we used the {\tt pamr} package, which implements NSC, to confirm that both our discovery and validation sets consisted of the correct samples and to reproduce their classifier for further NSC-LASSO comparisons. {\tt pamr} classification resulted in a test accuracy of 1.0, meaning it predicted every sample’s cluster correctly. This confirms that the discovery set, validation set, and the NSC classifier were successfully reproduced.

\subsection{LASSO Classification using {\tt glmnet}}
The {\tt glmnet} R package is used to fit the LASSO  and the Elastic Net. In our case, we fit a LASSO model, a generalized linear model with an L1 penalty, shrinking some coefficients and zeroing out others to create sparsity. LASSO uses a forward, discriminative method of classification, predicting a dataset's response variable $Y$ given the explanatory variable $X$ and discarding non-predictive features. The intensity of the sparsity penalty is determined by the $\lambda$ parameter, chosen using cross validation with the function ``cv.glmnet'' \citep{glmnet}. We perform cross validation on 10 reproducible, randomly selected folds and choose the lambda which minimizes the CV error, measured with misclassification error. In the function call, the $X$ training data is the genomic data in our discovery set and $Y$ is the  cluster labels in our discovery set. We set family to “multinomial”, which trains a multinomial classification model. We set foldid to 10 reproducible, randomly selected folds; we set type.multinomial either to “ungrouped” or “grouped”, depending on the type of classifier, explained briefly below; and we set type.measure to “class” such that our CV error is measured with a metric comparable to that of {\tt pamr}. We experimented with three different LASSO models: an ungrouped multinomial classifier, a grouped multinomial classifier, and a One-Versus-Rest (OVR) classifier.

\begin{figure}
    \centering
        \includegraphics[width=\textwidth]{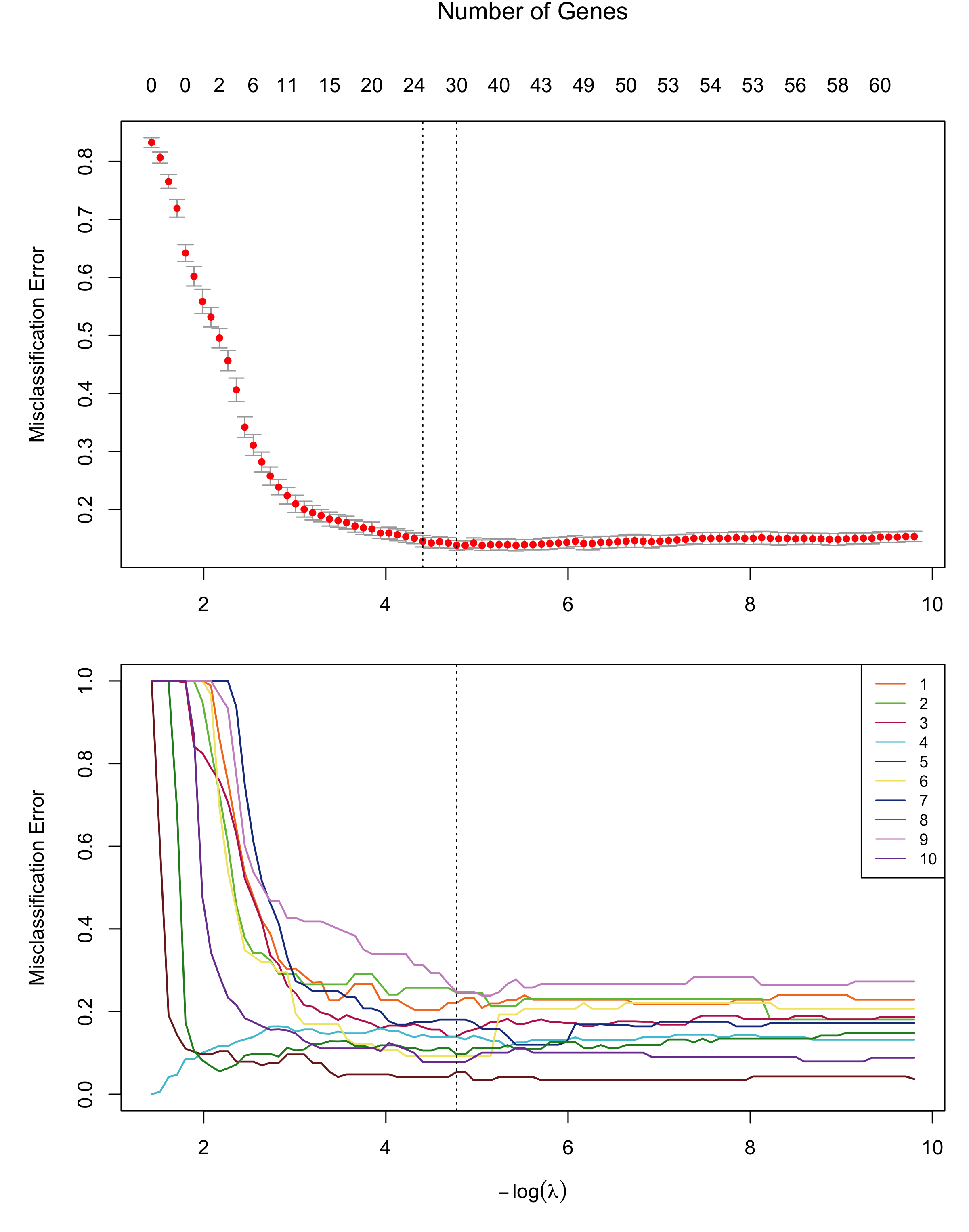}
        \caption{\textit{Ungrouped {\tt {\tt glmnet}} CV curve plot. From left to right, the two dotted lines and the bottom horizontal axis show the $\lambda$ value at which CV error is at its minimum (lambda.min = 0.00842, CV error = 0.138) and the maximum $\lambda$ value at which CV error is within one standard error from the minimum (lambda.1se = 0.0122, CV error = 0.145). The values at the top of each dotted line are the average numbers of features selected per class depending on the lambda value (number of features per class for lambda.min = 30). The bottom plot shows the CV curves for each cluster and the dotted line indicates the CV errors at ``lambda.min.''} }
        \label{fig2}
\end{figure}

The ungrouped multinomial classifier is the default LASSO multiclass model which penalizes each class’s coefficients for a variable separately, meaning that a variable's weight may be nonzero for some classes while zero for others. The grouped multinomial classifier penalizes all classes’ coefficients for a variable jointly, meaning that a variable is either nonzero or zero across all classes. Figure \ref{fig2} shows the CV curve plot from the ungrouped lasso model. 

In addition to these two multinomial models, we fit a One-Versus-Rest (OVR) LASSO model, which splits the multinomial classification task into several binomial classifiers, as many classes as there are in the dataset, which in our case is ten. Each classifier performs standard binomial classification: all rows equal to its respective class are set to ``1'' and the rest are zeroed out. Using the same folds as the previous two {\tt glmnet} models, the $\lambda$ value which minimizes CV error for each classifier is chosen. We then calculate a weighted average CV error by multiplying each classifier's minimum CV error by the number of patients in its corresponding cluster, summing the results, and dividing this sum by the total number of patients in the discovery set. We can compare this single value with the CV errors of the other {\tt glmnet} models and {\tt pamr}. 

\subsection{NSC Classification using {\tt pamr}}
In contrast to LASSO's forward, discriminative approach, NSC uses a backward, generative method of classification, predicting the explanatory variable $X$ given the response variable $Y$. The features within each class are assumed to have normal distributions with different centroids or means, but share a common covariance matrix. Restricting this covariance matrix to be diagonal leads to the NSC classifier. {\tt pamr} then uses soft thresholding for feature selection, which shrinks each class centroid's gene t-statistic to the overall mean or sets it to zero if it becomes negative \citep{THNC2002}. The larger the delta value, the stronger the shrinkage and the sparser the model.

\begin{figure}
    \centering
        \includegraphics[width=\textwidth]{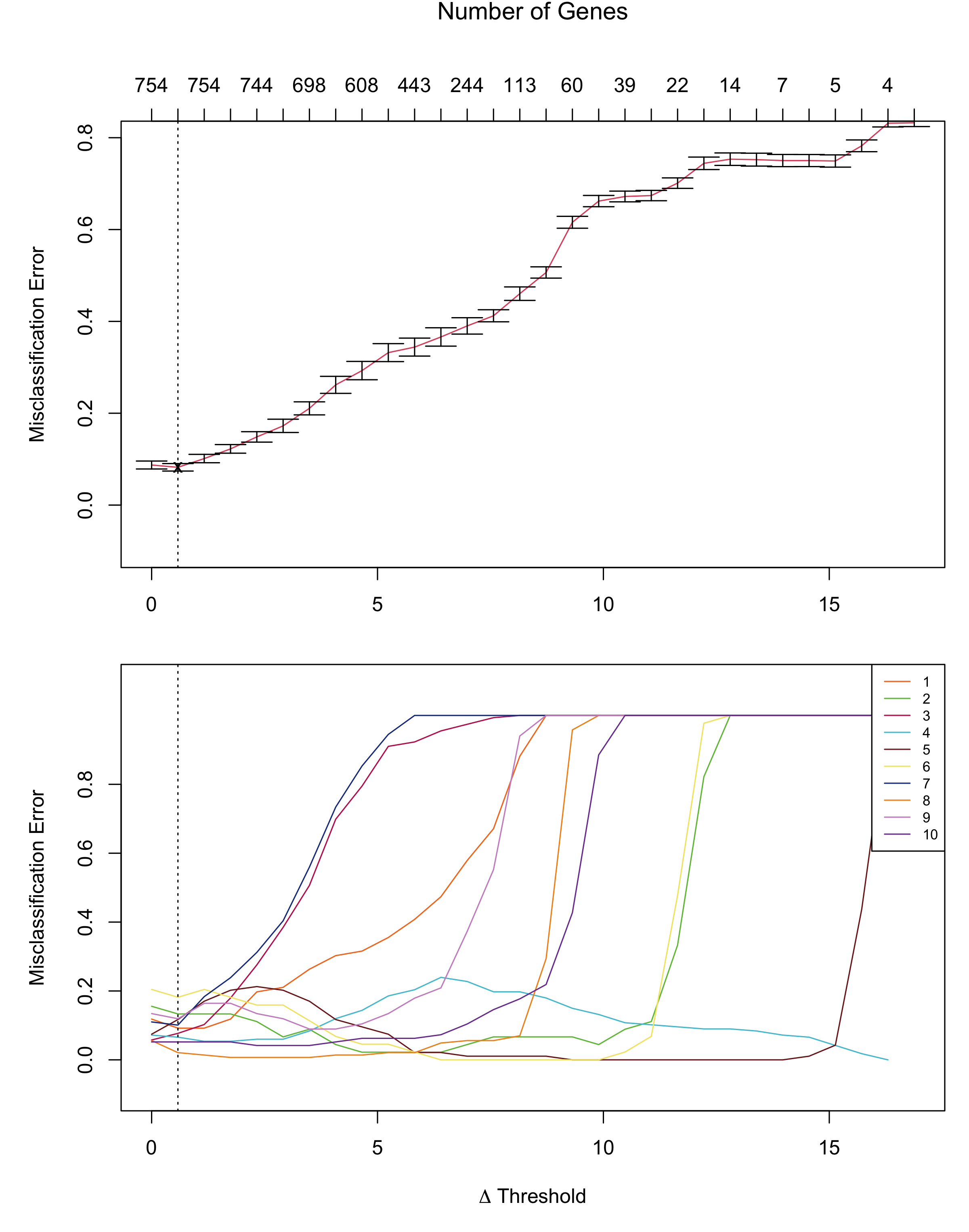}
        \caption{\textit{{\tt pamr} CV plot. The dotted line on the top plot shows the misclassification rate at the best delta threshold (best delta threshold = 0.582, CV error = 0.0822). The values at the top of each plot are the number of features or genes selected depending on the delta threshold value (number of features for best delta threshold = 754). The bottom plot shows the CV curves for each cluster and the dotted line indicates the CV errors at the best threshold.}}
        \label{fig3}
\end{figure}

To reproduce the NSC classifier used by \cite{curtis2012}, we called the ``pamr.train'' function which produces a sequence of thirty evenly spaced delta threshold values. We then performed cross validation on the same ten randomly created folds used in CV by the {\tt glmnet} models, fitting NSC on the held-out folds and measuring the CV error, or misclassification error, at each delta. It calculates the minimum CV error and records the first delta value in the sequence at which the CV error equals this minimum. We call this the ``best'' delta threshold. The {\tt pamr} CV curve plot is shown in Figure \ref{fig3}.

\begin{figure}
    \centering
        \includegraphics[width=\textwidth]{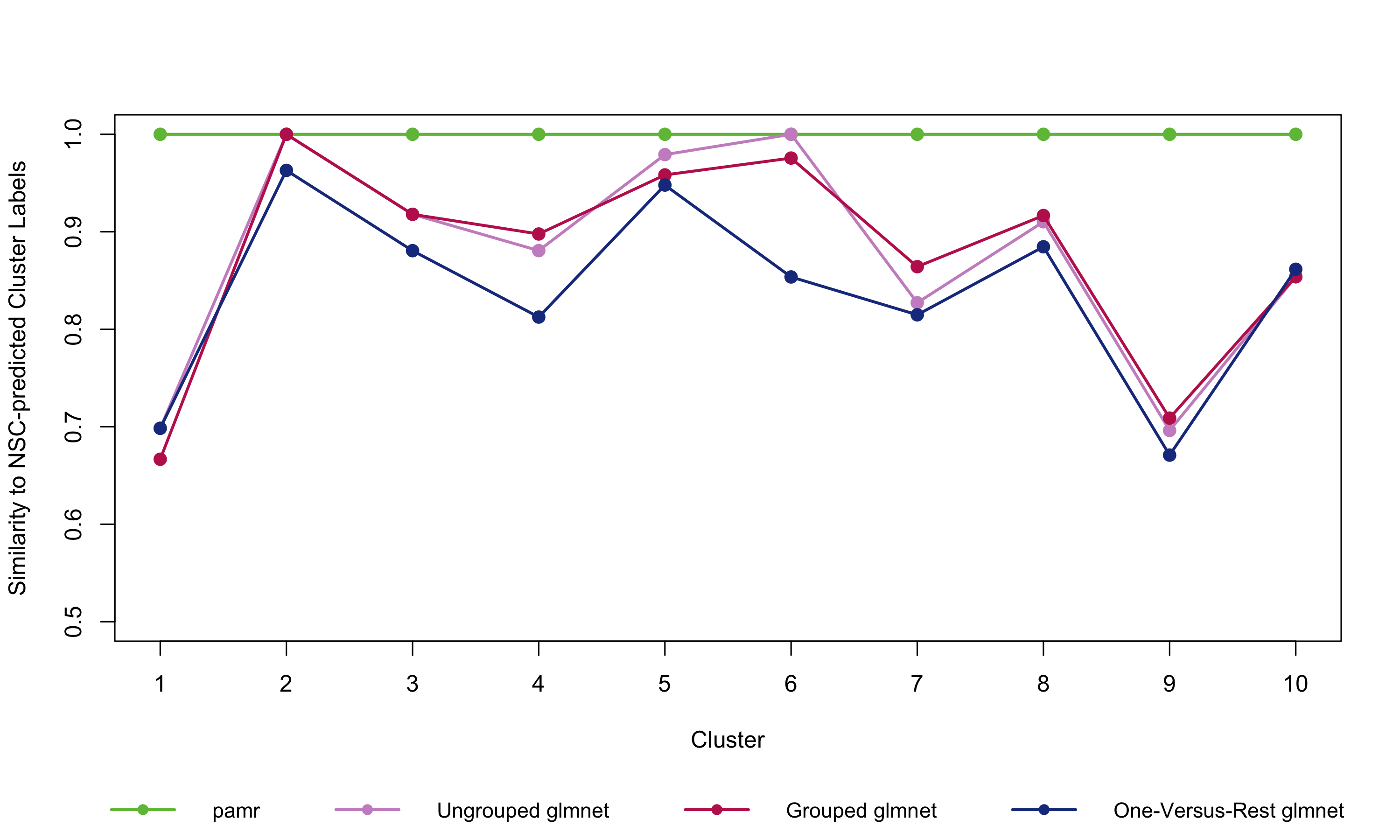}
        \caption{\textit{Similarity to NSC-predicted cluster labels plot. The similarity score is measured by the proportion of patients assigned to a cluster by both {\tt glmnet} and {\tt pamr} out of the patients assigned to it by {\tt pamr} only. The {\tt pamr} model scores 1.0 across all clusters since it reproduces the NSC classifier and thus predicts the same clusters for all patients as those in the downloaded cluster data. All other points show how similar the predicted clusters of the {\tt glmnet} models' are to the NSC-predicted labels. Out of all the LASSO models, the OVR model labels are the least similar to those of the NSC classifier as evidenced by its lowest scores for eight clusters.}}
        \label{fig4}
\end{figure}

To measure the similarity between the LASSO-predicted and NSC-predicted clusters of the validation set, we evaluated each {\tt glmnet} classifier against the labels of the validation set, which were downloaded with the rest of the data. For each group, we calculate the proportion of patients assigned to it by both {\tt glmnet} and {\tt pamr} out of the patients assigned to it by {\tt pamr} only. Since the {\tt pamr} model reproduces the NSC classifier, it yielded a score of 1.0 for all clusters. The scores of the three {\tt glmnet} models are shown in Figure \ref{fig4}.

\section{Model Comparison}
\subsection{CV Error and Sparsity}
\begin{figure}
    \centering
        \includegraphics[width=\textwidth]{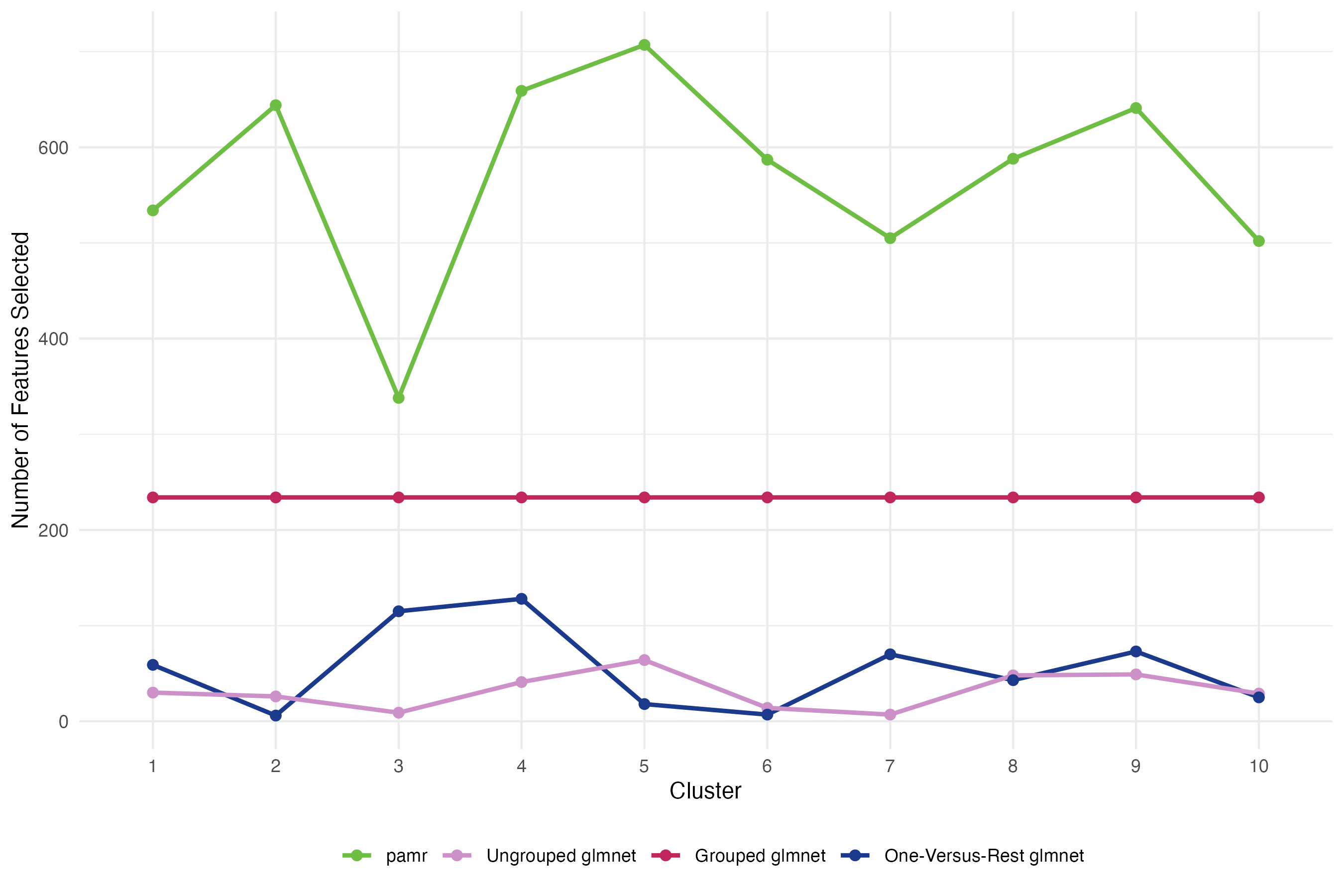}
        \caption{\textit{Number of features selected for each cluster.}}
        \label{fig5}
\end{figure}

{\tt glmnet}'s ungrouped multinomial model selects 267 total features, producing a minimum CV error of 0.138; the grouped multinomial model selects 234 total features, producing a minimum CV error of 0.106; the OVR model selects 401 total features, producing an average minimum CV error of 0.0572; and the {\tt pamr} model selects 754 total, or all, features, producing a CV error of 0.0822. Figure \ref{fig5} shows the number of features selected per model across all clusters. The OVR {\tt glmnet} model produces a lower CV error and a sparser model than the NSC classifier, suggesting that the One-Versus-Rest model is both a more accurate and more efficient classifier. However, the NSC classifier yields a lower CV error rate than both the default ungrouped and grouped {\tt glmnet} models. 

After each model is fit, we save the predicted class objects created by {\tt glmnet}'s predict function and {\tt pamr}'s ``pamr.predict'' function, to be compared to the ``true'' classes of the discovery set to evaluate the predictive accuracy of each model on the validation set. 

\subsection{Kaplan-Meier Curves}
\begin{figure}
    \centering
        \includegraphics[width=\textwidth]{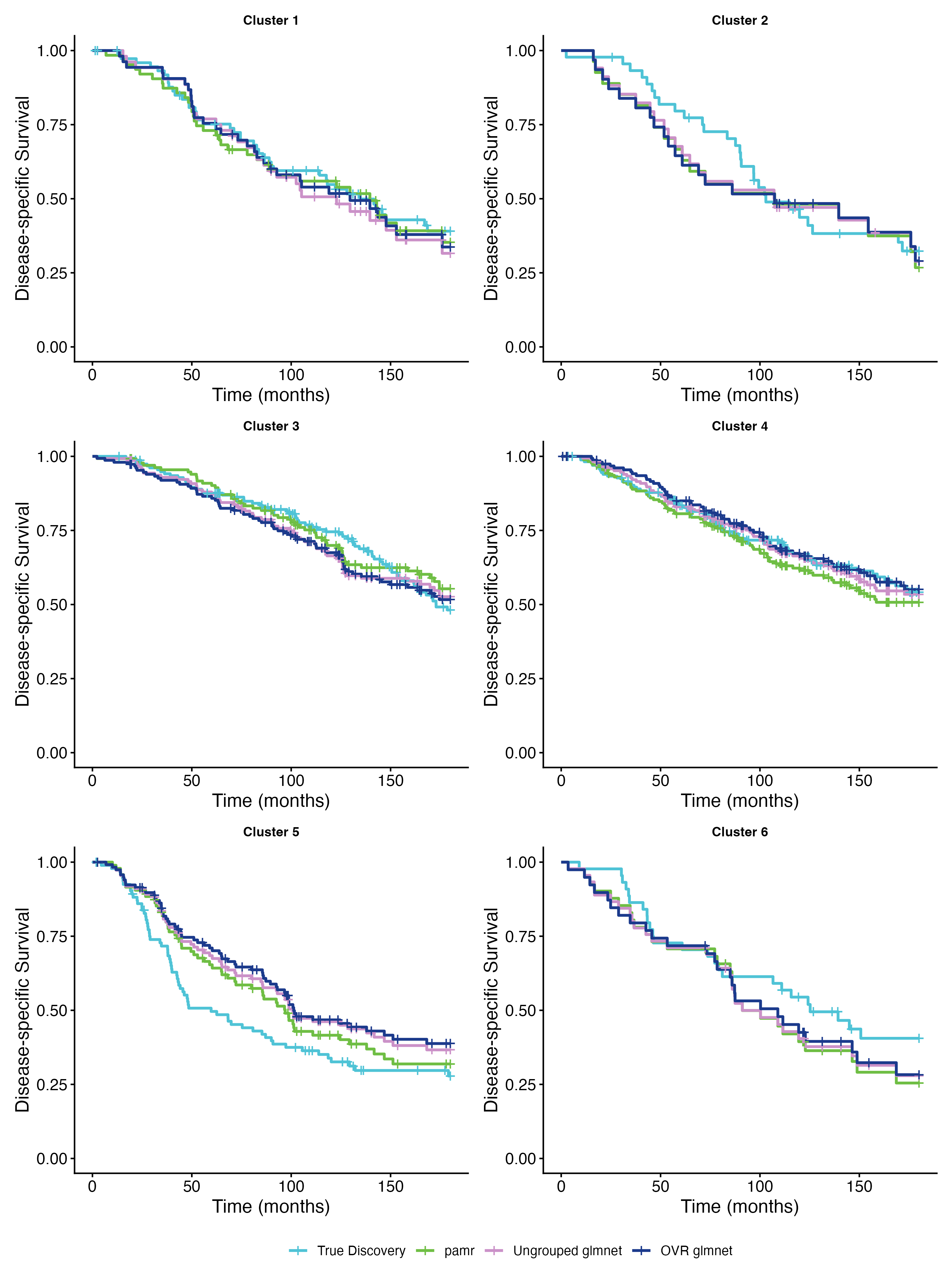}
        \caption{\textit{KM curves panel per cluster for clusters 1 to 6. The vertical axis plots the disease-specific survival probability of breast cancer patients in each cluster from 0 to 180 months based on their aggregate survival data.}}
        \label{fig6}
\end{figure}

\begin{figure}
    \centering
        \includegraphics[width=\textwidth]{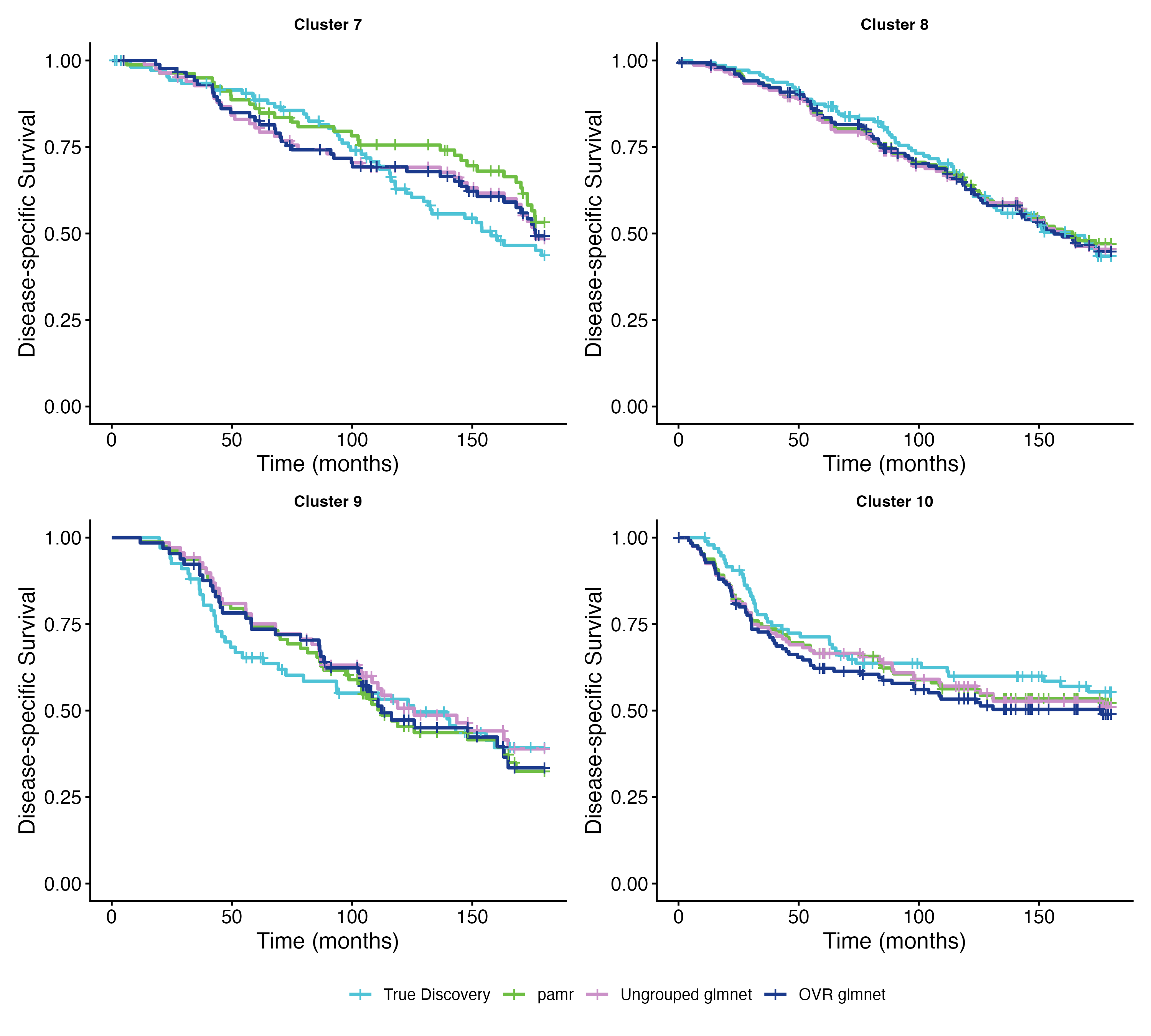}
        \caption{\textit{KM curves panel per cluster for clusters 7 to 10. The vertical axis plots the disease-specific survival probability of breast cancer patients in each cluster from 0 to 180 months based on their aggregate survival data.}}
        \label{fig7}
\end{figure}

We visualize the comparisons between the discovery set cluster labels and each model's predicted cluster labels with Kaplan-Meier (KM) curves. Using the survival R package, these curves show the survival trajectory of each cluster group of patients. We created a panel of 10 KM curve plots, one for each cluster, shown in Figures \ref{fig6} and \ref{fig7}. Within each plot is the true discovery, {\tt pamr}, and the ungrouped {\tt glmnet} KM curves. 

\begin{figure}
    \centering
        \includegraphics[width=\textwidth]{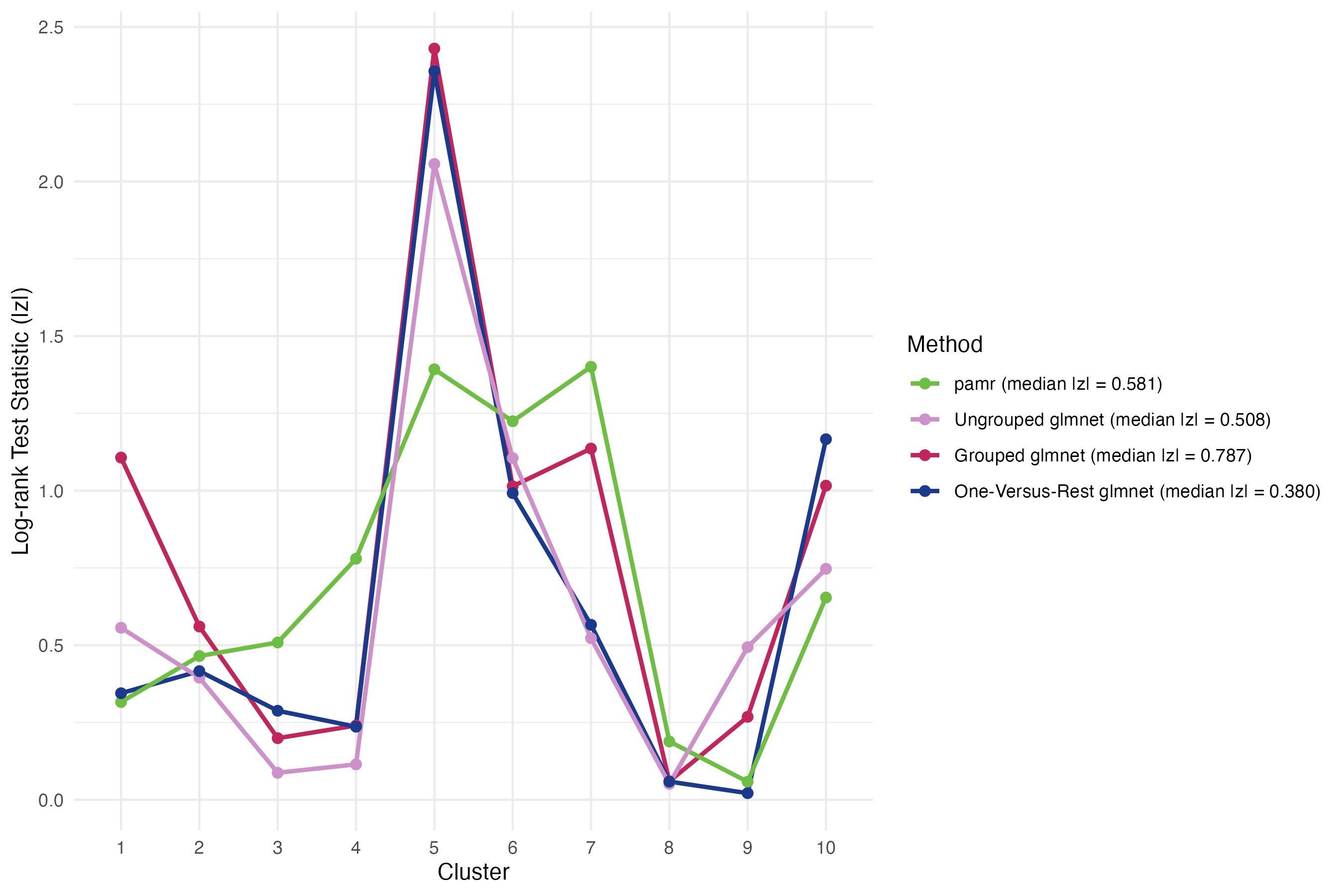}
        \caption{\textit{Log-rank test statistics per cluster for each model. For clusters 3, 4, 6, 7, and 8, all three {\tt glmnet} KM curves are more similar to the true discovery KM curves than are the {\tt pamr} model's. OVR {\tt glmnet}'s log-rank test are lower than {\tt pamr}'s for clusters 2, 3, 4, 6, 7, 8, and 9. For the rest, {\tt pamr} performs better than at least one {\tt glmnet} model. The median z-scores for ungrouped, grouped, OVR, and {\tt pamr} are 0.508, 0.787, 0.380, and 0.581, respectively.}}
        \label{fig8}
\end{figure}

In order to objectively compare the similarity between the true discovery KM curve and each model's KM curves, we calculate one log-rank test statistic, or the absolute value of the signed z-score, per model  per cluster, using the {\tt survdiff} object generated using the survival package. A graph visualizing all models' log-rank statistics across the clusters is shown in Figure \ref{fig8}. Since we are taking the absolute values of the z-scores, the lower the value, the more similar the KM curve of the model is to the true discovery KM curves. 

While the ungrouped and grouped {\tt glmnet} models had weaker classification accuracies, for each cluster at which {\tt glmnet} has a lower z-score than {\tt pamr}, {\tt glmnet}'s classification of patients into that cluster produces a survival distribution more statistically accurate to the true discovery's distribution than that from {\tt pamr}'s classification. For example, the ungrouped {\tt {\tt glmnet}} model produced KM curves more similar to the true discovery set's KM curves for six clusters. Since the ungrouped {\tt glmnet} CV error is higher than {\tt pamr}'s, then it can be said that the {\tt glmnet} curves' similarity to the true discovery curves cannot necessarily be attributed to {\tt glmnet}'s classification performance. 

Another {\tt glmnet} model which performed well with its log-rank test statistics in comparison to {\tt pamr} was the OVR model, which yielded a lower z-score for seven clusters. The median z-score was 0.380, the lowest out of all four models. The patients which OVR classified into each cluster have survival data more similar to those of the discovery set patients determined by integrative clustering.

\section{Discussion}
 
From the comparisons between the {\tt pamr} model and the three {\tt glmnet} models we applied to the METABRIC data, the LASSO multi-class classifiers offers competitive prediction accuracy for this problem in comparison to the Nearest Shrunken Centroids classifier, but they produced much sparser models. This sparsity can help in understanding the underlying biological mechanisms of the cancer classes. 

Among the LASSO models, the OVR model allows different choices of regularization per class by allowing different $\lambda$ values that minimize CV error for each cluster. Consistently, the OVR model selected fewer than half as many features as {\tt pamr} for each cluster. The OVR model also produced a weighted average CV error of 0.0572, the lowest out of all four models, showing that OVR's ten binomial classifiers perform better in predicting held-out folds of the discovery set than the other models. Moreover, according to its log-rank test statistics, OVR {\tt glmnet} most successfully reproduces in the validation set the survival distributions of each true cluster in the discovery set. The other two {\tt glmnet} models also offer sparser models, but sacrifice CV accuracy and, occasionally, similarity to the discovery set's KM curves. 

With a strong classifier and its consistent survival trajectory predictions, large datasets of patients can be divided into clusters such as these to infer disease-specific survival probabilities over time. When predictions are supported by supervised learning models such as Nearest Shrunken Centroids and LASSO, researchers are able to  better judge the treatments that are most appropriate for patients in each cluster. 

The code to run each classification model will be available on GitHub.

\bibliographystyle{agsm}
\bibliography{tibs}

test accuracy plot: start vertical axis at 0.5, add grouped and ovr, cluster labels under points

\end{document}